\documentclass{emulateapj}

\usepackage{longtable}
\usepackage{natbib}
\usepackage{graphicx}
\newcommand{\super}[1]{\scriptsize \ensuremath{^\textrm{#1}} \normalsize}

\begin{document}

\title{A near-infrared excess in the continuum of high-redshift galaxies: a tracer of
star formation and circumstellar disks?}

\author{\medskip 
Erin Mentuch\altaffilmark{1},
Roberto G. Abraham\altaffilmark{1},
Karl Glazebrook\altaffilmark{2}, 
Patrick J. McCarthy\altaffilmark{3}, 
Haojing Yan\altaffilmark{4}, 
Daniel V. O'Donnell\altaffilmark{5},
Damien Le Borgne\altaffilmark{6,7},
Sandra Savaglio\altaffilmark{8}, 
David Crampton\altaffilmark{9}, 
Richard Murowinski\altaffilmark{9}, 
St{\'e}phanie Juneau\altaffilmark{10}, 
R. G. Carlberg\altaffilmark{1},
Inger J{\o}rgensen\altaffilmark{11}, 
Kathy Roth\altaffilmark{11}, 
Hsiao-Wen Chen\altaffilmark{12}, and 
Ronald O. Marzke\altaffilmark{13}}

\altaffiltext{1}{Department of Astronomy \& Astrophysics, University of Toronto, 50 St. George Street, Toronto, ON, M5S~3H4, Canada}

\altaffiltext{3}{Observatories of the Carnegie Institution of Washington,813 Santa Barbara Street, Pasadena, CA 91101, USA}

\altaffiltext{2}{Centre for Astrophysics and Supercomputing, Swinburne University of Technology, 1 Alfred St, Hawthorn, Victoria 3122, Australia}

\altaffiltext{4}{Center for Cosmology and AstroParticle Physics, the Ohio State University, 191 West Woodruff Avenue, Columbus, OH 43210, USA}

\altaffiltext{5}{Department of Physics, Ernest Rutherford Physics Building, McGill University, 3600 rue University, Montreal, QC, H3A 2T8, C}

\altaffiltext{6}{Institut d'Astrophysique de Paris, UMR7095, UPMC, Paris, France}

\altaffiltext{7}{CNRS, UMR7095, Institut d'Astrophysique de Paris, F-75014, Paris, France}

\altaffiltext{8}{Max-Planck-Institut f\"ur extraterrestrische Physik, Garching, Germany}

\altaffiltext{9}{Herzberg Institute of Astrophysics, National Research Council, 5071 West Saanich Road, Victoria, British Columbia, V9E~2E7, Canada.} 

\altaffiltext{10}{Department of Astronomy/Steward Observatory,University of Arizona,933 N Cherry Ave., Rm. N204,Tucson AZ 85721-0065, USA}

\altaffiltext{11}{Gemini Observatory, Hilo, HI 96720, USA}

\altaffiltext{12}{The Department of Astonomy and Astrophysics, University of Chicago, 5640 S. Ellis Ave, Chicago, IL 60637, USA}

\altaffiltext{13}{Dept. of Physics and Astronomy, San Francisco State University, 1600 Holloway Avenue, San Francisco, CA 94132, USA}

\begin{abstract}

A broad continuum excess in the near-infrared, peaking in the rest-frame at 2-5\,\micron, is detected in a spectroscopic sample of 88 galaxies at $0.5<z<2.0$ taken from the Gemini Deep Deep Survey. Line emission from polycyclic aromatic hydrocarbons (PAHs) at 3.3\,\micron~alone cannot explain the excess, which can be fit by a spectral component consisting of a template of PAH emission lines superposed on a modified blackbody of temperature $T\sim850$\,K. The luminosity of this near-infrared excess emission at 3\,\micron~is found to be correlated with the star formation rate of the galaxy. The origin of the near-infrared excess is explored by examining similar excesses observed locally in massive star forming regions, reflection and planetary nebulae, post-asymptotic giant branch stars and in the galactic cirrus. We also consider the potential contribution from dust heated around low-luminosity active galactic nuclei. We conclude that the most likely explanation for the 2-5\,$\micron$ excess is the contribution from circumstellar disks around massive young stellar objects seen in the integrated light of high-redshift galaxies. Assuming circumstellar disks extend down to lower masses, as they do in our own Galaxy, the excess emission presents us with an exciting opportunity to measure the formation rate of planetary systems at cosmic epochs before our own Solar System formed.

\end{abstract}


\section{Introduction}

The spectral energy distributions (SEDs) of most normal star-forming 
galaxies can be well-described by a model which is simply the 
superposition of a component due to starlight and a component 
due to re-radiated dust emission. The physical details of the stellar component 
are well understood on the basis of many 
decades of observations made at visible wavelengths, but
it is only recently that near (NIR) and mid-infrared (MIR) 
space-based observatories have revealed the richer spectral 
information contained in the dust component.
This component is dominated by broadband thermal emission
from dust grains,
but emission
associated with
polycyclic aromatic hydrocarbons (PAHs) has also been
found to be important in the NIR and MIR \citep{des90}. 
 Sensitive spectroscopic and broadband observations from the 
 \textit{Spitzer Space Telescope} \citep{wer04} and \textit{Infrared Space Observatory} \citep{kes96} of both the
 PAH emission and the warm thermal component, combined with laboratory measurements
of the physical properties of interstellar dust provide a general understanding
of dust in the context of galactic properties (see for example \citealt{dra07}) and
can provide unique clues to our understanding of prominent dust processes such as metal enrichment and embedded star formation.

In large surveys of high-z galaxies, broadband photometry is commonly fit to SEDs produced by stellar population synthesis models to determine photometric redshifts, stellar masses, star formation rates (SFRs) and other global quantities of galaxies. Although, the high number of parameters involved in these models leads to large degeneracies in the fitted parameters, stellar masses can be measured robustly if rest-frame $K$-band light is known. However, at longer wavelengths the contribution from PAHs and emission from other non-stellar sources of luminosity are not included in the models. For this reason, band fluxes beyond rest-frame 2\,\micron\ are often excluded from stellar population synthesis SED fitting routines (see for example \citealt{pap06,per08,mag08}). A more ideal approach is to include non-stellar
sources of emission in the SED models to help break the degeneracies that
plague stellar population synthesis models, such as those associated with age, metallicity, dust attenuation.

A new result that has emerged recently is the discovery of
an excess in the NIR continuum of galaxies at 
2-5\,\micron, \citep{lu03,hel04,mag08}. This excess 
cannot be attributed solely to narrow emission
  from the 3.3\,\micron~PAH feature and is instead
  due to a combination of line emission and an additional thermal component. 
  The emission can be significant at wavelengths around 3\micron, but it
  is not a major contributor to a galaxy's total bolometric flux. For example,
  in a survey of 45 star-forming 
  galaxies \cite{lu03} measure a weak excess 
  component of continuum emission in the range of 2-5\,\micron~
  that is characterized by a very high color temperature ($\sim1000$\,K), but
  is only a few percent of the far-infrared (FIR; 8-1000\,\micron) luminosity 
  attributed to larger dust grains. 
The origin and physical mechanism of the NIR excess in galaxies is 
not known, although \cite{lu03} suggest it likely originates in the ISM.

An NIR excess continuum has also been seen locally in a range of objects. Reflection nebulae (RNae) around massive O and B-type stars have a smooth continuum feature from 1.25 to 4.8\,\micron~that is characterized by a blackbody with a $\sim1000$\,K temperature in addition to a narrow emission feature at 3.3\,\micron~\citep{sel83,sel96}, attributed to the thermal emission of very small ($\sim10$\,\AA) grains \citep{sel84}. $JHKL$ photometry of massive stars in the $\sim1-3$\,Myr old star forming regions NGC\,3576 \citep{mae06}, G305.2+0.2 \citep{lon07}, and 30 Dor in the LMC \citep{mae05} show L-band (3.5\,\micron) excesses in more than half of their stars with masses $>$ 10\,$M_\odot$ that is attributed to circumstellar disks \citep{lad92}. Excess continuum emission can also come from more diffuse sources. Observations by \textit{Spitzer's} Infrared Array Camera (IRAC; \citealt{ faz04}) and the {\it Cosmic Background Explorer}/Diffuse 
Infrared Background Explorer \textit{COBE/}DIRBE of the galactic cirrus \citep{fla06} reveal a continuum component in the NIR whose contribution to the IRAC\,[3.6 \micron]-band is $50\%-80$\%, while the PAH feature at 3.3\,\micron~contributes the remaining flux. 

\begin{figure*}[tbp]
\plotone{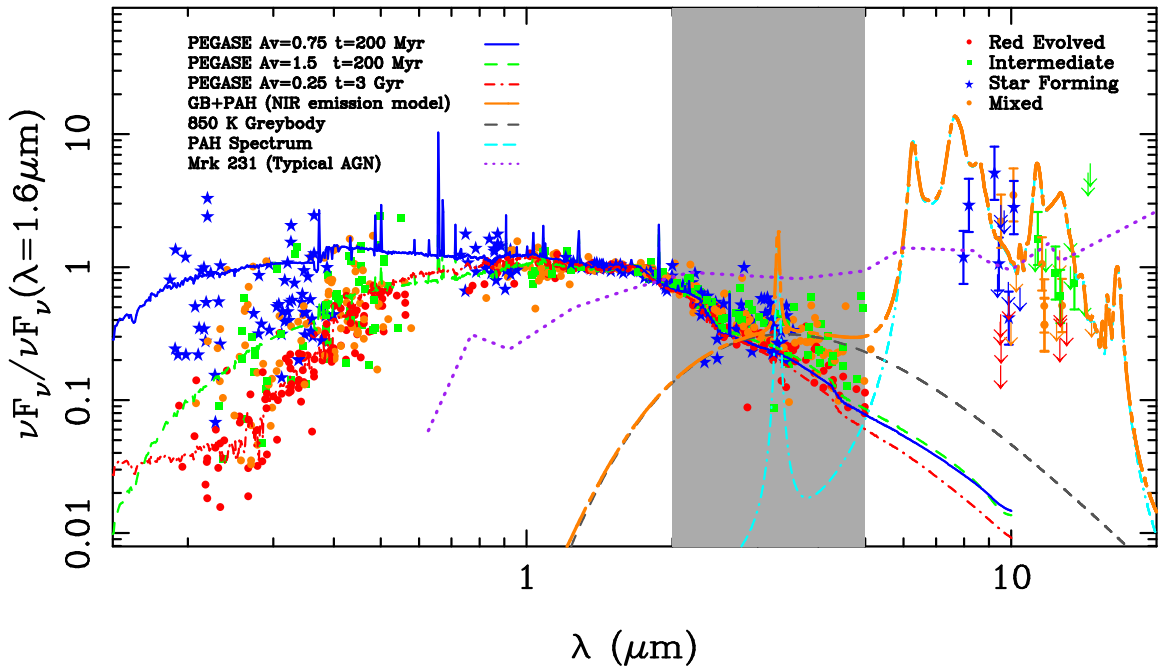}
\caption{VIzK$_{s}$+IRAC and MIPS photometry from the GDDS normalized at $\lambda=1.6\,\micron$ with the 2-5\,\micron~region of NIR excess highlighted in gray. Plot symbols are keyed to galaxy spectral type, as shown in the legend and described in the text (Section \ref{s:spec}). The blue solid curve is a stellar component model from PEGASE.2 \citep{fioc97} for a typical star-forming galaxy (t=200 Myr, A$_{\mathrm{v}}$=0.75). A dusty star-forming galaxy (t=200\,Myr, A$_{\mathrm{v}}$=1.5) and evolved galaxy (t=3\,Gyr; A$_{\mathrm{v}}$=0.25) are also plotted. The orange dash-long-dashed line represents the NIR emission component model (GB+PAH) from \citet{dac08} that is supplemented in our SED modeling. An 850\,K graybody emitter (gray dashed line) contributes most of the emission of this component at 2-5\,\micron, although the 3.3\,\micron~PAH feature (the cyan dot-dashed curve) has some contribution ($<20$\,\% in a given IRAC band). For reference, an SED of an AGN (Mrk 231; purple dotted curve) is plotted. For pedagogical purposes, the GB+PAH and AGN curves have been normalized to the average MIPS flux of our sample. For those objects with MIPS detections, the AGN model is not a good representation of the photometry. (A color version of this figure is available in the online journal.) 
}
\label{fig:sed}
\end{figure*}

The ultimate cause of $\sim 1000$\,K blackbody emission in the empirical 
data described above is still unknown. Large dust grains are fragile at 
such high temperatures, and are unlikely to be the source of the excess emission. 
In the canonical interstellar dust model of \citet{des90}, PAH molecules, unlike
dust grains, can exhibit both line
emission at 3.3\,\micron~ and broad continuum emission at 2-5\,\micron, 
and lead to significant NIR emission.
Large PAH molecules ($>$1\,nm) display strong broadband 
NIR emission and are more resilient to 
strong ionizing fields, while smaller PAH molecules display a larger 
line-to-continuum emission ratio in the NIR. This would naturally lead
to broad NIR continuum emission of molecules in the vicinity of
massive stars, likely contained in circumstellar disks, a mechanism for 
the NIR excess in galaxies not
previously explored. Theoretically for local young stellar objects (YSOs),
 \citet{dul01} show that the puffed up inner edges
of circumstellar disks can radiate at high temperatures and 
lead to L-band excess luminosities in the NIR of more than $1000\,L_\odot$
 in the most massive stars.  In agreement with the circumstellar disk 
 interpretation of the excess, observations presented in \citet{are94}
  with the \textit{COBE/}DIRBE show that NIR
continuum emission of nearby star-forming regions correlate well 
with bright mid-IR observations and that the dust must lie very close 
to the stars, as it was unresolved by the DIRBE beam. More recently, 
resolved excess NIR continuum emission is observed in young stars 
that possess circumstellar disks (see Figure~3 in \citealt{woo08}).

Whatever the cause of the emission, we will show in this paper that
it is apparently fairly easy to detect such excess emission at high redshifts, at least in an optically 
selected sample like the Gemini Deep Deep Survey (GDDS; \citealt{bob04}). At first glance this seems rather
surprising, since (as we have already noted) the excess emission corresponds to only a tiny fraction of a typical galaxy's
bolometric flux.  However, as demonstrated in Figure~\ref{fig:sed}, the peak wavelength 
of a blackbody radiator emitting at $\sim 1000$\,K (gray dashed curve) falls near 
the global minimum in a typical galaxy's SED when modeled as a
two-component system of starlight  + IR emission. Therefore the contrast
of even a small amount of  $\sim 1000$ K  emission seen at 2-5\,\micron~ can be significant,
giving us an opportunity to probe a new component of a galaxy's flux that is ordinarily overwhelmed
by either starlight or re-radiated emission from large dust grains. This new
component is interesting on its own and also may be useful for calibration of the SFR.
Studies of the NIR component in ultra-luminous infrared galaxies (ULIRGS; \citealt{mou90,ima06,mag08}) suggest that the 3.3\,\micron~emission can be used as a good indicator of the SFR in some galaxies, although further calibration and study of star-forming galaxies is needed to justify this as a useful diagnostic. 

In this paper, we quantify the NIR excess continuum in a spectroscopic sample of high redshift galaxies, described in Section \ref{s:obs} by performing linear least-squares fits of broadband optical+IRAC photometry to SED models. We compare two sets of SED models, the first consisting of a stellar component SED only and the second, a two component SED model consisting of both a stellar component and an additional NIR dust emission component consisting of an 850\,K graybody plus PAH emission template \citep{dac08} as described in Section \ref{s:methods}. The case for an NIR excess continuum in GDDS galaxies is made through IRAC color-color diagrams and SED example fits in Section \ref{s:res}. We show that the luminosity of the NIR emission component is correlated with a galaxy's star-formation rate in Section \ref{s:sfr}. In Section \ref{s:dis} we speculate on the origin and potential significance of the excess, and show that the most likely candidate is a population of heated circumstellar disks in the high redshift galaxies.

Throughout this paper, we
adopt a concordance cosmology with \hbox{$H_0$=70 km s$^{-1}$ Mpc$^{-1}$},
$\Omega_M=0.3$, and $\Omega_\Lambda=0.7$. All photometric magnitudes are given
in AB magnitudes, unless otherwise noted.

\section{Observations}\label{s:obs}

\subsection{Summary of GDDS Observations}\label{s:spec}

The Gemini Deep Deep Survey (hereafter GDDS; \citealt{bob04}) is a spectroscopically defined color selected subset of the 1 deg\super{2} Las Campanas Infrared Survey (LCIRS; \citealt{mcc02}). It consists of four  5\arcmin.5 $\times$ 5\arcmin.5 fields chosen from the parent sample to minimize cosmic variance. The sample consists of galaxies with $K_{s,VEGA} < 20.6$ mag and $I_{VEGA} < 24.5$ mag ($K_{s,AB} < 22.4$ mag and $I_{AB} < 24.9$ mag) in the redshift interval $0.5 < z < 2.0$. The GDDS multi-band coverage includes $UVBRIz'JHK_{s}$ broadband colors in addition to high-resolution Advanced Camera for Surveys (ACS) coverage of $\sim60$\% of the fields. For this analysis, we only select objects with high-confidence spectroscopic redshifts and those which are confidently detected in the longest IRAC channel (we required $\Delta [8.0\,\micron] < 0.3\,$mag). This results in 103 galaxies from the 309 objects in the original GDDS sample. 

All objects were given a spectral type classification as outlined in Table\,5 of \cite{bob04}. The classification is based on three digits that refer to young, intermediate-age, and old stellar populations. Objects showing nearly pure signatures of an evolved stellar population were assigned a GDDS class of ``001" in \cite{bob04}. In Figure~\ref{fig:sed} and many figures in this paper, we key each galaxy to their galactic spectral type. The evolved ``001'' galaxies are plotted as red circles. Objects showing a strong-UV continuum dominated by massive stars were classified as ``100" and are plotted as blue stars.  Objects with intermediate ages (e.g., strong Balmer absorption) were classified as ``010" and are plotted as green squares. Some objects consist of both an evolved population plus a younger more recent star formation component. These objects are listed as mixed populations with spectral types of ``110" or ``101"  in  and are plotted as orange dots in our figures. 

\subsection{{\it Spitzer} IRAC Photometry}\label{s:obsIRAC}

Near-infrared observations are important for determining stellar masses through SED fitting, particularly at high redshift where the light from evolved stellar populations, which trace the stellar mass of a galaxy, is redshifted out of the optical. In order to determine stellar masses of GDDS galaxies, a total of 10.9 hours of \textit{Spitzer} IRAC observations were obtained for three of the four GDDS fields (Program ID: 3554; \citealt{mcc04}) yielding 5$\sigma$ depths of 0.45, 0.9, 6 and 8 $\mu$Jy in respective IRAC bands. The final field, GDDS-SA22, was observed through an existing GTO-reserved program (Program ID: 30328; \citealt{faz06}). Mosaics were created from the archived Basic Calibration Data (BCD) images using MOPEX (Final Version 18.2), a more current and improved post-BCD mosaicing tool of the \textit{Spitzer} Science Center to achieve better background removal and a spatial resolution of 0.6\arcsec/pixel. The first three short-exposure BCD images were thrown away to correct for the IRAC ``first-frame effect''. The background was corrected, for each remaining BCD image, by subtracting the peak of a count histogram of the entire image. The resulting [5.8\,\micron] and [8.0\,\micron] images displayed noticeable gradients and in addition had to be gradient-corrected. This gradient correction was not performed in the deeper GDDS-SA22 field. The resulting BCD images were background corrected to an AB magnitude zeropoint of 21.85 and drizzle-combined to create the final post-BCD mosaics.

Objects in the post-BCD IRAC mosaics were detected by running SExtractor in dual-image mode, where a weighted-sum of [3.6\,\micron] and [4.5\,\micron] images were used as the detection image. A  Mexhat filter was used instead of a Gaussian filter to better de-blend and detect objects in these relatively low resolution, but deep, images. Magnitudes were then measured in each IRAC band in 4\arcsec~ diameter apertures. To extrapolate the total flux of the galaxy, aperture corrections of 0.317, 0.360, 0.545 and 0.677 mags were added to the [3.6\,\micron], [4.5\,\micron], [5.8\,\micron], [8.0\,\micron] magnitudes, respectively.

Source blending from objects on the edge of our 4\arcsec~apertures could potentially add a flux excess. We went through our sample of 103 galaxies by eye (inspecting both our ground-based $I$-band and \textit{HST} ACS\,[814W] images) and flagged any object that has a neighbor within a 7\arcsec~aperture. We found 15 such objects in total and will show in Section 4 that these objects are relatively poorly fit by our SED models as a result of the photometric confusion, so we opt to exclude them from most analyses. Thus our final sample is comprised of
88 spectroscopically confirmed galaxies. This sample covers the full GDDS redshift range of $0.5<z<2$, with objects ranging from $z_\mathrm{spec}=0.57$ to $z_\mathrm{spec}=1.97$ (32 objects from $0.5<z<1$, 41 objects from $1<z<1.5$, and 14 objects from $1.5<z<2$).

\begin{figure}[t]
\begin{center}
\plotone{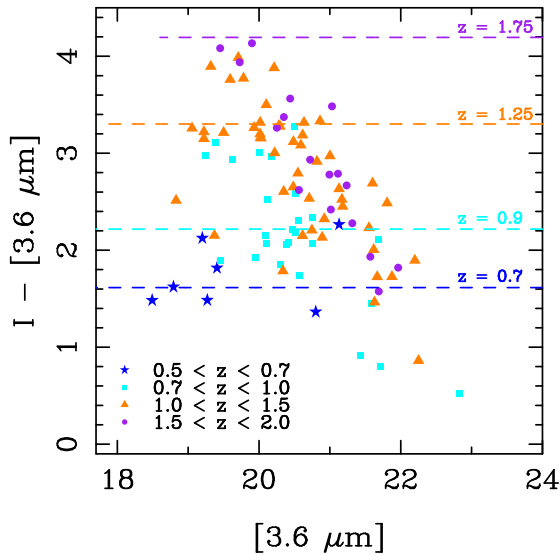}
\caption{$I-[3.6 \micron]$ vs. [3.6 \micron] color-magnitude (in AB magnitude system) plot. The distribution of GDDS galaxies in this parameter space is redshift dependent. We plot for reference the $I-[3.6\,\micron]$ color of a 3\,Gyr galaxy ($\tau=500$\,Myr, A$_{\mathrm{v}}$=0.25, Z=Z$_\odot$) at each redshift interval. (A color version of this figure is available in the online journal.) }
\label{fig:IRACopticalcolour}
\end{center}
\end{figure}

The $I-[3.6\,\micron]$ - [3.6\,\micron] color-magnitude diagram presented in Figure~\ref{fig:IRACopticalcolour} shows that $I$-[3.6\,\micron] color is a strong function of redshift for our mass-selected sample of galaxies. This dependence is most pronounced for the brighter galaxies, suggesting that for these galaxies the red sequence is set in place by $z\sim1.5$. We plot for reference the color of an evolved galaxy (age=3\,Gyr, $\tau=500$\,Myr, A$_{\mathrm{v}}$=0.25, Z=Z$_\odot$) at each redshift interval.  The brightest galaxies are well matched by this evolved population model, while the fainter galaxies are bluer, indicative of younger ages. 

\begin{figure}[t]
\begin{center}
\plotone{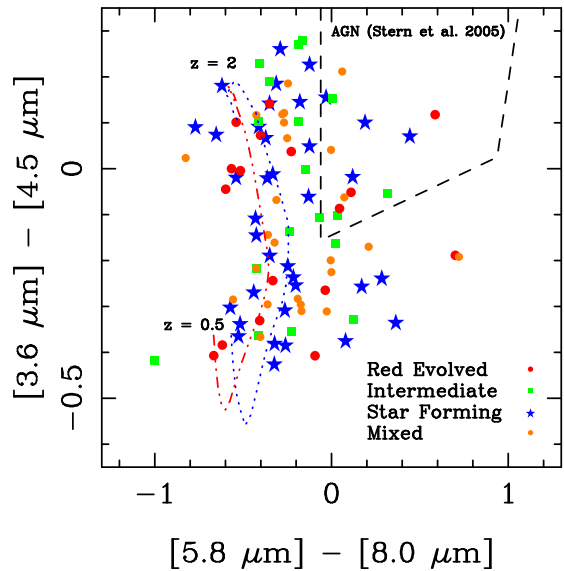}
\caption{IRAC color-color plot of [3.6\,\micron] - [4.5\,\micron] vs. [5.8\,\micron] - [8.0\,\micron] (in AB magnitude system). We plot for reference the evolution of a typical star-forming galaxy (blue dotted line; age=200\,Myr galaxy, $\tau=500$\,Myr, A$_{\mathrm{v}}$=1, Z=Z$_\odot$) and evolved galaxy (red dot-dashed line; age=3\,Gyr, $\tau=500$\,Myr, A$_{\mathrm{v}}$=0.25, Z=Z$_\odot$). AGN selection from \citet{ste05} (region contained by the dashed line) shows that the sample could contain a couple of AGNs according to these criteria. However, MIPS photometry further rules out contamination from strong AGN as seen in Figure~\ref{fig:IRACMIPS}. (A color version of this figure is available in the online journal.) 
}
\label{fig:IRACcolour}
\end{center}
\end{figure}

In Figure~\ref{fig:IRACcolour} an IRAC color-color plot for the sample is presented. For reference we plot the evolution of a star-forming and evolved galaxy computed with PEGASE.2 code. The GDDS sample generally traces these tracks, although many galaxies show redder [5.8\,\micron]-[8.0\,\micron] colors, possibly indicating an NIR excess at rest-frame 3\,\micron. We plot the IRAC color-color selection criteria to pick out an active galactic nucleus (AGN) from \cite{ste05} as the region contained in the dashed region in Figure~\ref{fig:IRACcolour}. Only 11\% of the objects from the selected GDDS sample fall in this region and most of those are near the selection boundary, showing little contamination from strong AGNs in our sample. Objects are plotted with plot symbols keyed to their galaxy spectral types using the scheme described in Section \ref{s:spec}. There is no obvious relationship between a galaxy's location in the color-color space and its spectral type.

\subsection{\it{Spitzer} 24\,\micron\ MIPS Photometry}\label{s:mips}

For one of our fields, GDDS-SA22, a 1320\,s exposure was observed with \textit{Spitzer's} Multiband Imaging Photometer for {\it Spitzer} (MIPS; \citealt{rie04}) under a GTO reserved program (Program ID: 30328; \citealt{faz06}). Our 24\,\micron~MIPS imaging was reduced using a combination of the \textit{Spitzer} Science Center's MOPEX post-processing suite and a set of custom routines developed to better remove background and artifacts. The source finding and photometry/extraction pipeline (D.V. O'Donnell et al. 2010, in preparation) employed PPP \citep{yee91}, DAOPHOT \citep{ste87} and additional routines for calibration. A Monte Carlo technique in which the theoretical MIPS point-spread function (PSF) was scaled, inserted into the science images and retrieved by the photometry pipeline was used to determine an aperture correction for all sources on the order of $\sim17$\% of the source's measured band flux.

Because a number of our objects had poor MIPS correlations, we inspected each source by eye to confirm whether they were in fact counterparts. Source confusion is a problem in some of our objects which we later flag in our analysis (Section \ref{s:sedresults1}). To any objects non-detected we assigned an upper limit of 70\,$\mu$Jy,
our 1\,$\sigma$ detection limit computed from Monte Carlo completeness tests. 

\begin{figure}[tbp]
\begin{center}
\plotone{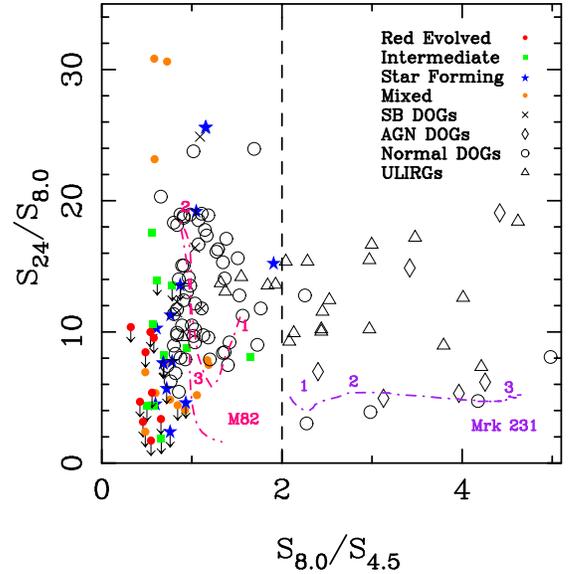}
\caption{IRAC-MIPS 24\,$\mu$m color-color plot for objects (classified by galactic spectral types as defined in Section \ref{s:spec}) in the GDDS-SA22 field. Non-detected MIPS objects are plotted as upper limits. Evolutionary tracks for a typical AGN (Mrk 231) and starburst (M82) (taken from \citealt{pop08a}) are shown for z=1,2,3,4. Objects on the left of the dashed line are classified as starburst dominated, and those on the right of the dashed line are AGN dominated \citep{pop08b}. ULIRGS from \citet{saj07} are plotted as open triangles, AGN-dominated, starburst-dominated and normal dust obscured galaxies (DOGs) at $z\sim2$ as classified in \citet{pop08b} are plotted as open diamonds, crosses, and open circles, respectively. (A color version of this figure is available in the online journal.) }
\label{fig:IRACMIPS}
\end{center}
\end{figure}

MIPS fluxes are available for 40/88 galaxies in our selected sample. The IRAC-MIPS 24\,$\mu$m color-color plot, presented in Figure~\ref{fig:IRACMIPS}, shows the majority of objects in the GDDS-SA22 field lie in the region of starburst galaxies, left of the dashed line, as defined by \citet{pop08b}, with a general trend of the star-forming and mixed population galaxies showing stronger S$_{24}$/S$_{8}$ ratios than the evolved galaxies. Non-detections (25/40) are plotted as upper limits. None of our objects have S$_{8.0}$/S$_{4.5}>2$, an indication of a significant AGN component \citep{ivi04,pop08b}. We are thus confident that our optically and mass-selected sample does not have a large contamination of strong AGN. Due to selection biases between our sample and other high-z AGN samples, we cannot completely rule out contamination due
to lower luminosity AGN (see later discussion in Section \ref{s:agn}).

For comparison with other high-z galaxy samples, we also show in Figure~\ref{fig:IRACMIPS} the locations in the IRAC-MIPS color-color plane of a sample of 24\,\micron-selected ULIRGS at $z\sim1-3$ \citep{saj07} and a sample of faint 24\,\micron-selected  $z\sim2$ dust obscured galaxies (DOGs; \citealt{pop08a}). \citet{pop08b} use X-ray, far-IR, sub-mm and radio observations to determine whether the DOGs have strong AGN or starburst components which we plot as separate symbols in Figure~\ref{fig:IRACMIPS}. No GDDS galaxy shows similar IRAC-MIPS colors as the class of AGN-dominated galaxies. For reference, evolutionary tracks for a typical AGN (Mrk 231) and starburst (M82) (taken from \citealt{pop08a}) are shown. It is evident that the GDDS mass-selected sample probes the less active  parameter space of high-z galaxies (i.e., lower SFRs, smaller contributions from the AGN).

\section{Spectral Energy Distribution Fitting}\label{s:methods}

We aim to understand the NIR excess by incorporating emission 
from dust as an additional component in each galaxy's SED.
 Each galaxy's stellar component is modeled 
using the techniques of population synthesis. In contrast to the 
well-worn techniques of stellar population synthesis, our modeling of 
dust is rather simplistic. However, we will show that even relatively 
crude dust models can provide interesting constraints. As 
astronomical facilities which explore faint galaxy populations 
continue to probe further into the IR (e.g., JWST, ALMA) it will 
become increasingly important to embrace dust as an essential 
component when modeling galaxy SEDs. 
Our analysis might therefore be viewed as an early attempt to 
move in this general direction.

\subsection{Stellar component}\label{s:onecomp}

To model the stellar SED component, a library of SEDs is generated consisting of a range of  two component star formation histories (SFHs). We generate two sets of SED libraries, one from PEGASE.2 software \citep{fioc97} and one from \citet{mar05} (hereafter M05), whose treatment of thermally pulsing asymptotic giant branch (TP-AGB) stars is scaled to match photometry of globular clusters in the NIR. Both sets consist of a primary SFH component consisting of a monotonically evolving stellar population with a SFR $\propto \exp(-t/\tau)$, $\tau \epsilon \left[0.1,500\right]$\,Gyr for PEGASE.2 models and $\tau \epsilon \left[0.1,20\right]$\,Gyr for M05 models. The first approximates an instantaneous starburst and the last a constant SFR. A secondary SFH component models a starburst event, with a $\tau=0.1$~Gyr exponential. In the PEGASE.2 models, this secondary component can occur at any time between $z_{form}$ and $z_{obs}$, while in the M05 models it is fixed to 50\,Myr before $z_{obs}$ since the SFH is not a user input parameter in the M05 models, as it is in the PEGASE.2 models. We note here that fixing this parameter does not affect the relevant parameters ($A_v$, $M_{\star}$) derived from the SED fits. The strength of the second burst component can have a mass between $10^{-4}$ and 2 times the initial component. The two component stellar model allows for a more robust determination of stellar masses by correctly modeling the light from evolved populations and that from a recent star forming episode. We opt to use the initial mass function (IMF) from \citet{kro01}.

Extinction is added to the spectra ourselves in eleven reddening bins covering $0\le A_{\mathrm{V}}\le2$\,mag. We assume that nebular extinction is twice the stellar continuum and use the Small Magellanic Cloud (SMC) reddening law of \citet{pei92} which provides a numerical fit to the extinction in the SMC. For the PEGASE.2 models, evolution of the stellar population assumes constant metallicity, and does not include effects due to galactic winds, infall or substellar objects. The PEGASE.2 models vary in five possible metallicity values from $0.0004 \leq Z \leq 0.02$, while the M05 models are fixed to solar metallicity.

We refer the reader to Figure~\ref{fig:sed} for examples of the SED model components plotted with the normalized (at $\lambda = 1.6\,\micron$) rest-frame fluxes of our data set. The observed NIR excess region is highlighted in gray. Two examples of a young (t=200\,Myr), star-forming galaxy are shown with different extinction values as the blue solid ($A_{\mathrm{v}} =0.75$) and green dashed ($A_{\mathrm{v}}=1.5$) curves. An evolved stellar population (t=3\,Gyr; $A_{\mathrm{v}}=0.25$) is plotted as the red dot-dashed curve. These \textit{illustrative} SEDs, computed with PEGASE.2, are for a galaxy with an exponentially declining SFR of $\tau$=500\,Myr and $Z=Z_\odot$. 

\subsection{Additional dust component}\label{s:twocomp}

Broadband multi-wavelength (UV, optical, and IR) empirical SED modeling performed by \citet{dac08} of galaxies from the \textit{Spitzer} Infrared Nearby Galaxies Survey (SINGS; \citealt{ken03,dal05}) show that emission in the NIR, from 2-5\,\micron, can be modeled by an additional SED component described by a PAH template spectrum superposed on an 850\,K graybody modified by $\lambda^{-1}$ extinction (hereafter written as GB+PAH to identify this as a single parameter SED component). The PAH template is scaled to match the NIR continuum excess (e.g., the graybody) seen in star-forming galaxies in \cite{lu03} \& \cite{dal05} and is shown to also match the relative, but not absolute, intensities of the galactic cirrus (see Figure~1 of \citealt{dac08}). For reference we plot the GB+PAH component as the orange dash-long-dashed line in Figure~\ref{fig:sed}. As well, we show the separate components that comprise this model: an 850\,K graybody (dark gray dashed line), and a PAH template spectrum (cyan dot-dashed line). 

\subsection{Methods}

First we fit optical $VIz'K_{s}$ [3.6\,\micron] [4.5\,\micron] photometry to the stellar component models (both PEGASE.2 and M05). \footnote{To justify that the longer IRAC bands are 
inadequately modeled by stellar component only models, we also fit all the $VIz'K_{s}$ [3.6\,\micron] [4.5\,\micron] [5.8\,\micron] [8.0\,\micron] photometry. This also allows us to compare the quality of fits (see {s:qof}) between the single stellar component model and the two component stellar plus NIR emission model.} We then fit all optical $VIz'K_{s}$ + IRAC\,[3.6\,\micron]\,[4.5\,\micron]\,[5.8\,\micron]\,[8.0\,\micron] photometry to the second model set consisting of the same library of stellar component SEDs with the additional GB+PAH component just described in Section \ref{s:twocomp} fit as a sum of linear least-squares. The normalization of this component is thus an additional free parameter. 

A grid of apparent AB magnitudes is generated by redshifting each rest-frame model to the galaxy's spectroscopic redshift and convolving this spectrum with the transmission curves for all observed filters: $VIz'K_{s}$ and \textit{Spitzer} IRAC [3.6\,\micron], [4.5\,\micron], [5.8\,\micron], [8.0\,\micron]. 
The $\chi^{2}$ for each model is calculated in flux space with the normalization of each component as a fitted parameter. We restrict the age of the model to be less than the age of the Universe at the observed redshift. To estimate errors in the fitted parameters, we consider 20 Monte-Carlo realizations of the data through a random Poisson distribution of the count rate, thus simulating the detector shot noise.

\section{Results}\label{s:res}

\subsection{SED fitting: the stellar component model}\label{s:sedresults1}

\begin{figure}[tbp]
\begin{center}
\plotone{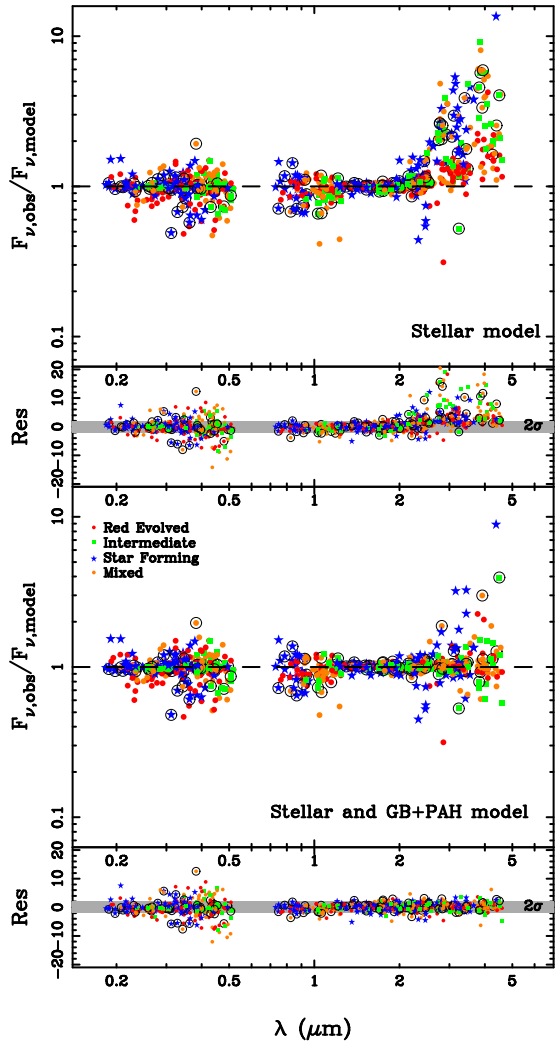}
\caption{Ratios of the observed band fluxes to the best-fitting SED band fluxes for each galaxy in the sample, each blueshifted to the rest-frame wavelength. The residual between observed and model band fluxes are given as ($F_{obs}-F_{model})/\sigma$ under each panel. In both plots, we flag objects with possible contamination in the IRAC channels due to source confusion as thin open circles. \textit{Top}: the optical (VIz'K) and \textit{Spitzer}/IRAC [3.6 \micron] and [4.0 \micron] bands are fit to a stellar only SED model.  \textit{Bottom}: an additional GB+PAH component is added to the SED model as described in Section \ref{s:methods} and SEDs are fit again using all VIz'K+IRAC bands (with reasonable detections). The two component SED model fits the data much better, except for the objects with known contamination from source blending in the IRAC channels (thin open circles, Section \ref{s:mips}). We exclude these objects in all subsequent analyses and plots. (A color version of this figure is available in the online journal.) }
\label{fig:restframeexcess}
\end{center}
\end{figure}

Figure~\ref{fig:restframeexcess} shows the observed-to-model flux ratios and residuals of the broadband optical+IRAC photometry fit to the two sets of models as described in Section \ref{s:methods}; one consisting of a stellar component only and the other consisting of two components, a stellar and an 850\,K GB+PAH SED component. Here, stellar models from M05 are shown, but the results are consistent with residuals from fits to the PEGASE.2 models. The wavelength has been blue-shifted to the rest-frame wavelength for comparison. The top panel shows the fit of the stellar component to the optical $VIzK_{s}$+IRAC\,[3.6\,\micron]\,[4.5\,\micron] photometry. \textit{Spitzer} IRAC [5.8 \micron] and [8.0 \micron] band fluxes are also compared to the model band fluxes, although they were not used to find the best-fit SEDs. We separate the data points into their optical spectral types as described in Section \ref{s:spec}.  For the $V$-[4.5 \micron] bands, the residuals are on average $0.08\sigma$ with a scatter of $2.34\sigma$, while the residuals for the [5.8 \micron] and [8.0 \micron] bands have a significant offset of $3.99\sigma$ and a scatter of $4.53\sigma$. Thus the stellar component models do a good job estimating the light blueward of 2\micron~where the flux is driven by the evolved population, and as a consequence galaxy stellar mass. However, redward of 2\,\micron, there is a significant excess in the majority of GDDS galaxies seen in the rest-frame range of 2-5\,\micron. 

We flag points in this figure (using thin circles) to identify galaxies that are possibly contaminated by light from nearby galaxies due to blending of the IRAC PSF (see Section \ref{s:obsIRAC}). On the whole, these systems are poorly fit by the models, and will drop them from subsequent figures and analyses, although including them changes none of our conclusions. 

A single or two-component SFH stellar population model alone is not sufficient to describe the flux generated in the NIR beyond rest-frame 2\,\micron. The amount of NIR excess is dependent on the galaxy spectral type as we see strong excesses of $F_{obs}/F_{model} = 3.3 \pm 2.5$ in the star-forming galaxies (blue stars), and $F_{obs}/F_{model} = 3.2 \pm 2.0$ in the mixed and intermediate populations (green squares and orange dots) and a smaller, but still noticeable excess of $F_{obs}/F_{model} = 1.5 \pm 0.6$ in galaxies dominated by an evolved stellar population (red circles). The NIR excess covers a broad spectral range indicating that a single narrow emission line such as the 3.3\,\micron~PAH emission line is not sufficient to explain all of the excess. 

\begin{figure*}[tbp]
\begin{center}
\plotone{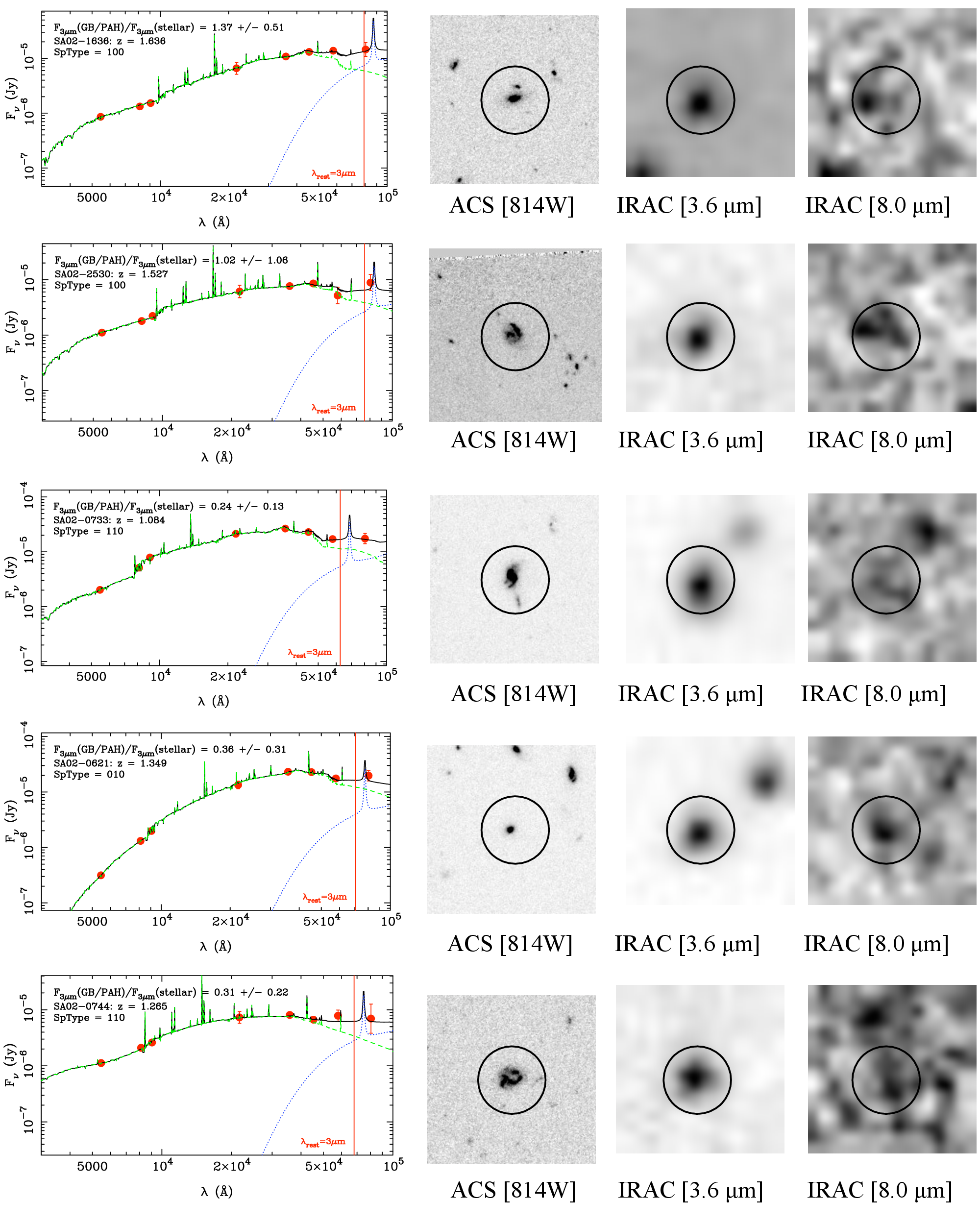}
\caption{SED fit examples for objects (band fluxes shown as red dots) with SFRs $>1\,M_\odot$/yr. The SED (solid black line) is composed of both a stellar component (green dashed) and an NIR emission component consisting a 850\,K graybody emission and a PAH spectrum (blue dotted line). 10\arcsec postage stamps from {\it Hubble's} ACS [814W], IRAC [3.6\,\micron] and [8.0\,\micron] are shown for each object. The spectral type (see Section \ref{s:spec}) of the galaxy is given in the top left corner showing the excess is seen in star-forming, intermediate, and mixed population galaxies. For reference, $\lambda_\mathrm{rest} = 3$\,\micron~is plotted as a red vertical line. The black circles show the 4\arcsec\ apertures used to derive our photometry. (A color version of this figure is available in the online journal.) }
\label{fig:SEDexamples}
\end{center}
\end{figure*}

\begin{figure*}[tbp]
\begin{center}
\plotone{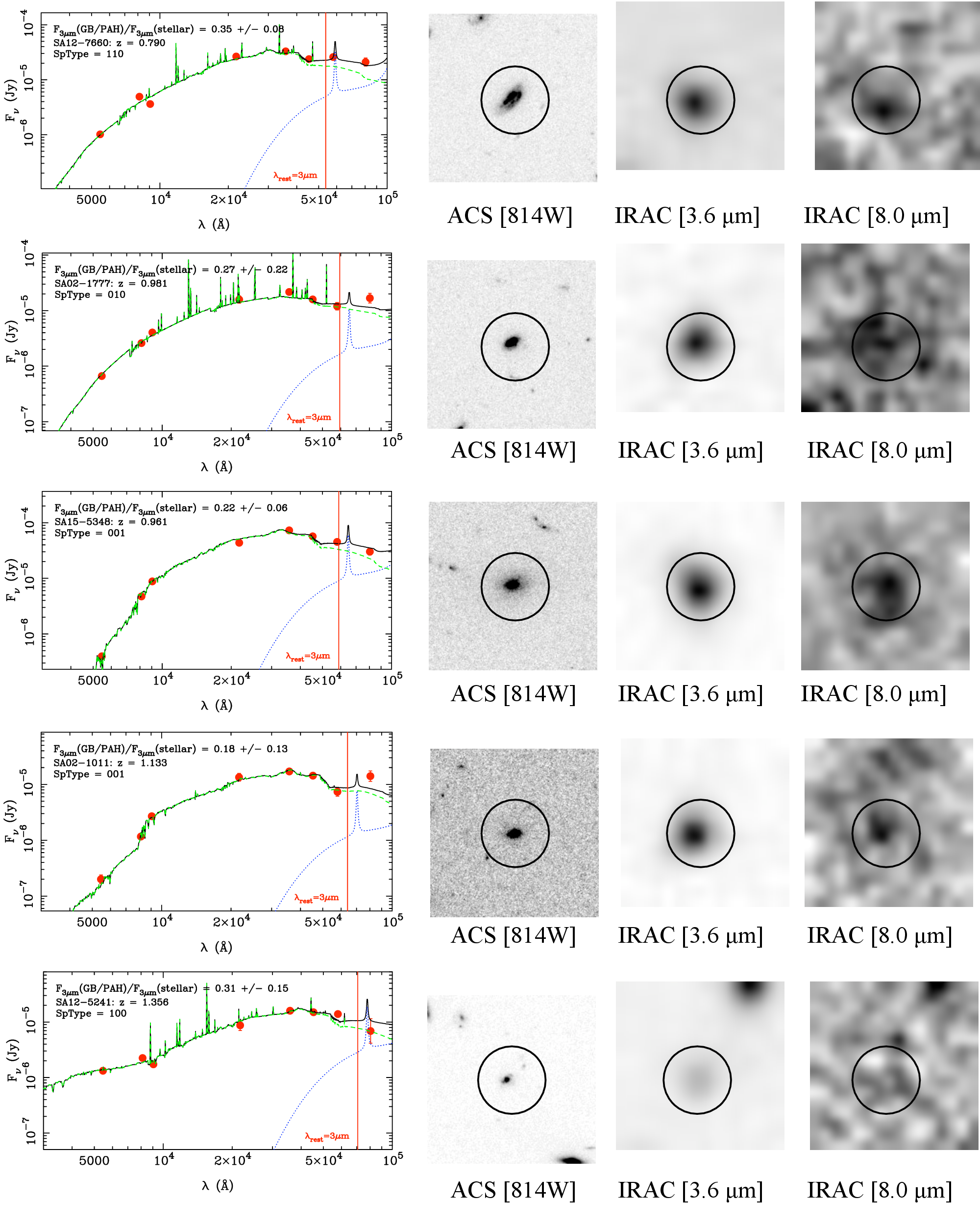}
\caption{As in Figure~\ref{fig:SEDexamples}, but for objects (band fluxes shown as red dots) with SFRs $<1\,M_\odot$/yr. (A color version of this figure is available in the online journal.) }
\label{fig:SEDexamples2}
\end{center}
\end{figure*}

\subsection{SED fitting: additional dust component}\label{s:sedresults2}

The second set of models consists of two components: one, a stellar SED component and the other, an NIR emission component consisting of 850\,K graybody emission plus a PAH spectrum (GB+PAH) that is scaled with the 850\,K graybody to match NIR observations of star-forming galaxies and the galactic cirrus  \citep{dac08}. The resulting ratio and residuals between observed and model band fluxes are plotted in the bottom panel of Figure~\ref{fig:restframeexcess} with wavelength blue-shifted to the rest-frame. The additional component matches the IRAC photometry, particularly the [5.8\,\micron] and [8.0\,\micron] bands, remarkably well.  For the $V$-[4.5 \micron] bands, the residuals are similar to the fits with just a stellar component model with a negligible offset of $0.07\sigma$ and a scatter of $2.24\sigma$, while the residuals for the [5.8 \micron] and [8.0 \micron] bands no longer have a significant offset with an average value of $0.25\sigma$ and a much smaller scatter of $1.58\sigma$.

Examples of fitting the two component SED to the optical+IRAC photometry are presented in Figures~\ref{fig:SEDexamples} and \ref{fig:SEDexamples2}. The black solid line represents the sum of the stellar (dashed green line)  and the additional GB+PAH component (blue dotted line). We plot 10\,\arcsec\ postage stamps from \textit{HST} ACS images of the galaxy to show morphology of each galaxy. There is no strong correlation between the amount of NIR excess and galaxy morphology, both qualitatively or quantitatively according to the asymmetry or Gini coefficient morphological quantities as measured for the GDDS sample in \cite{bob07}. We also plot images from IRAC [3.6\,\micron] and [8.0\,\micron], and show the 4\arcsec~apertures used for our photometry.

\subsection{Quality of fits}\label{s:qof}

Comparing the top and bottom panels of Figure~\ref{fig:restframeexcess}, we see
that the addition of a GB+PAH component greatly improves the fits qualitatively. 
The SED fitting routine minimizes $\chi^{2}$ but we emphasize that the
interpretation of the absolute value of the minimum $\chi^{2}_\mathrm{reduced}$ returned requires some careful consideration,
since the number of fitted parameters exceeds the number of data points.
This is a very common occurrence in broadband photometry SED fitting (e.g., \citealt{gla04}),
the consequence of which is that output model parameters are expected to be strongly correlated, and
interpretation of statistical significance is mainly based on Monte Carlo simulations. 
We calculate errors and assess the statistical significance of fit improvements 
by refitting the photometry to randomized versions of the single best-fit SED model.

We compare $\chi^{2}_\mathrm{reduced}$ values from least-square fits with 
both sets of models. We fit the stellar component model two times, once
with the inclusion of the two longer IRAC channels, and once with the [5.8\,\micron] 
and [8.0\,\micron] bands excluded. We then fit the optical and all IRAC channels 
to the two component model. The lowest quality SED fits are when all 
IRAC channels are fit to a stellar component SED model (median 
$\chi^{2}_\mathrm{reduced} = 2.69$). The highest quality fits are to the two 
component model (median $\chi^{2}_\mathrm{reduced}=1.73$), which are just
slightly better than the fits to the stellar model excluding the two
longer IRAC channels (median $\chi^{2}_\mathrm{reduced}=1.96$).
The additional component improves the ensemble quality of fit by 
 $\sim50\%$, which our Monte Carlo simulations indicate is far above the nominal
improvement expected from additional free 
parameters when fitting to an underlying SED with no excess flux. Thus, an SED  modeled by both a stellar component and a near-infrared
emission dust component represented by a 850\,K GB+PAH emission spectrum is a significantly better model of a galaxy's near-infrared light than one based on stellar emission alone.

\subsection{Graybody temperatures and other limitations on the dust model}

The choice of NIR emission model is based on observations compiled by \citet{dac08}. Our observations do not constrain any of the parameters in the model, such as the relative scaling between the NIR continuum and PAH emission or the temperature of the continuum emission.  Due to our limited spectral coverage, we only allow its normalization to be a free parameter. Other graybody temperatures were explored and we find that the continuum excess is equally fit (yielding consistent $\chi^{2}$ values) by graybody temperatures ranging from 700 to 1500\,K. As a comparison, \citet{fla06} find that a $1100\pm300$\,K works well for the galactic diffuse medium, while \citet{lu03} show that both a $750$\,K and a $1000$\,K graybody with $\lambda^{-2}$ emissivity can match the IRAC photometry. A different extinction law ($\lambda^{-2}$) was also explored but provided no improvement in the quality of fit. 

We attempted to add the temperature of the graybody as an additional parameter but found it was not possible to constrain the blackbody temperature at lower temperatures once the spectral peak of the graybody shifted outside the range of our  observed photometry (for reference, an 850\,K blackbody peaks at about 3.4\,\micron, which our IRAC observations straddle throughout $0.5<z<2$). Finally, fits to a model consisting solely of the 3.3\,\micron~PAH emission feature do
not succeed, showing only modest improvements to $\chi^{2}_\mathrm{reduced}$ and 
unrealistically high equivalent widths,
as the broad wings of the line are forced to fit the broad continuum excess.
The ratio of the PAH line emission to continuum emission could not be
constrained by our broadband observations, but follow up with NIR spectroscopy may
prove fruitful in constraining neutral to ionized PAH fractions and PAH size distributions (e.g., \citealt{des90,dra07}).

\subsection{Model consistency with MIPS photometry}\label{s:resultsMIPS}

\begin{figure*}[t]
\begin{center}
\plotone{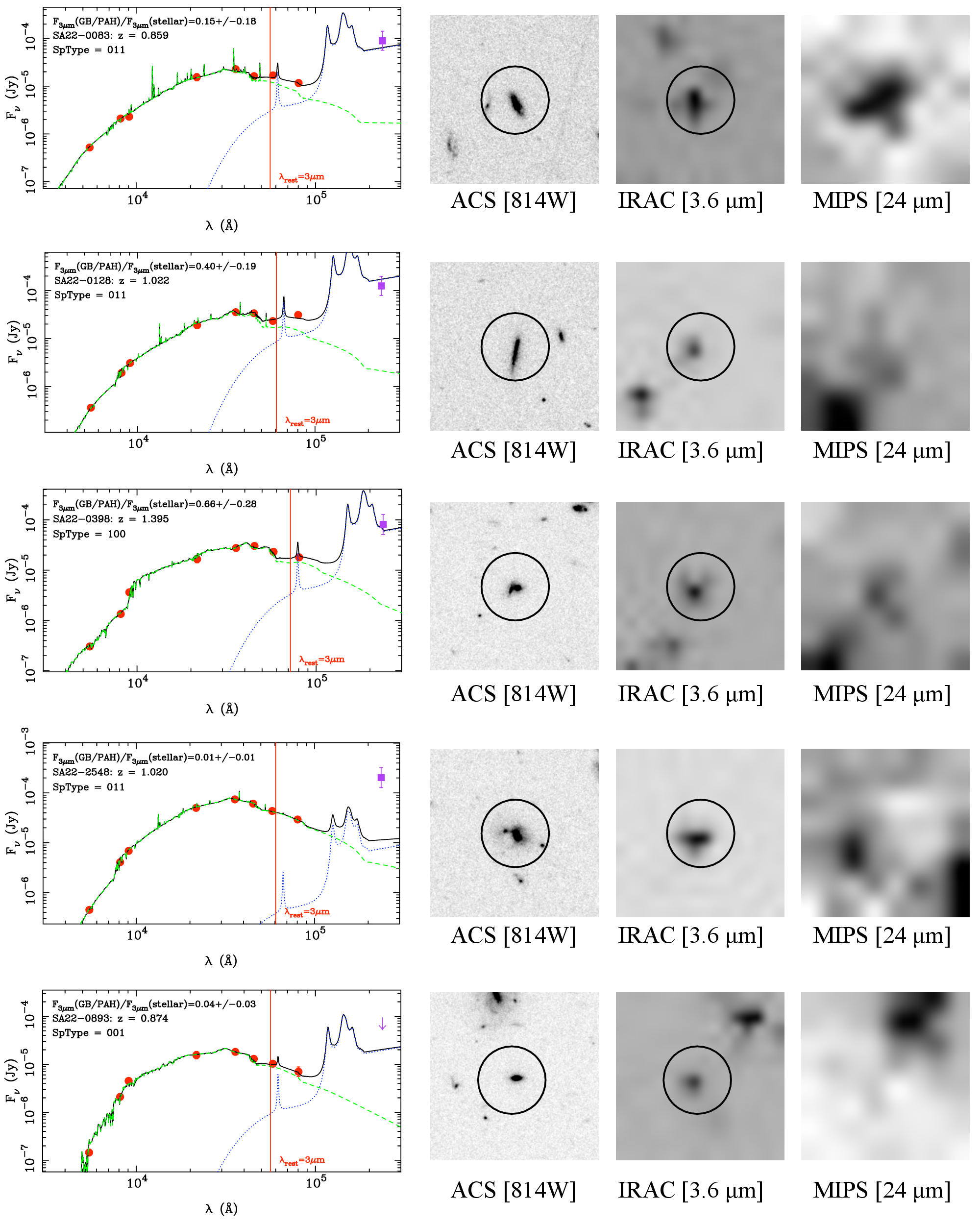}
\caption{SED fit examples for objects with MIPS detections are plotted in the first three rows. The measured MIPS fluxes (purple squares) are in agreement with those fluxes predicted by the stellar and GB+PAH two component model. ACS [814W], IRAC [3.6\,\micron] and MIPS [24\,\micron] postage stamps, of 10\,\arcsec~wide, are shown with the IRAC and optical apertures. We do not plot apertures on the MIPS figure which are measured as described in Section \ref{s:mips}. SED lines are the same as in Figure~\ref{fig:SEDexamples}, except now the wavelength range has been extended to include the MIPS data point. (A color version of this figure is available in the online journal.)  }
\label{fig:SEDexamples:MIPS}
\end{center}
\end{figure*}

Almost half (40/88) of our subset of GDDS objects with spectroscopic redshifts are found in the GDDS-SA22 field,
for which archival MIPS data are available. 
Since MIPS data were not incorporated into any of
our fits, any post-facto agreement between the MIPS
observations and our model strongly supports the central
idea of the NIR excess continuum being scaled to the MIR PAH spectrum. The MIPS 
band probes rest-frame 8-12\,\micron, where PAH features are known to exist. 
Thus we visually compare the MIPS band
fluxes with the GB+PAH SED component (orange dash-long-dashed line in Figure~\ref{fig:sed}), using the normalization of this component derived from our two component spectral
fitting. All non-detected (25/40) objects in the GDDS-SA22 field
are consistent with the predicted 24\,\micron\ fluxes ($<70\,\mu$Jy) from the GB+PAH component.  Only 15 out of the 40 objects are detected at 24\,\micron\ with MIPS. Eleven of these 15 MIPS-detected objects show 24\,\micron\ fluxes consistent with the best-fit SED model predictions. 
To illustrate the degree of
concordance, in Figure~\ref{fig:SEDexamples:MIPS} we show typical fits for a range of galaxy spectral types. The figure shows 
NIR excesses for 3 of the 11 objects with confident 24\,\micron~detections, as well as an example showing
a typically informative non-detection (last row in the figure). Only four 
galaxies do not have 24\,\micron~fluxes consistent with the model SEDs,
and three of these
show obvious source blending in the \textit{HST} ACS images
as a result of the larger MIPS PSF.

Therefore only a single galaxy (SA22-2548; plotted in the 4th row of Figure~\ref{fig:SEDexamples:MIPS}) is not consistent with the stellar and GB+PAH two component model
for that object (computed entirely independently of MIPS)\footnote{SA22-2548 is unusual in other ways: it has strong 24\,\micron~flux slightly offset from the optical GDDS counterpart, and the ACS image shows some indication of interaction,
so its 24\,\micron~excess may be the result of a strong AGN due to merging (although its IRAC colors ([3.6\,\micron] - [4.5\,\micron] = -0.216; [5.8\,\micron]-[8.0\,\micron] = -0.426) do not seem consistent with this).}. 
We conclude that an
additional GB+PAH model is consistent with the measured 24\,\micron~ MIPS fluxes for 97.5\% of the GDDS objects observed in the only GDDS field with MIPS data. 

\subsection{Correlation of the Near-Infrared excess with the Star Formation Rate}\label{s:sfr}

\begin{figure*}[thbp]
\begin{center}
\plottwo{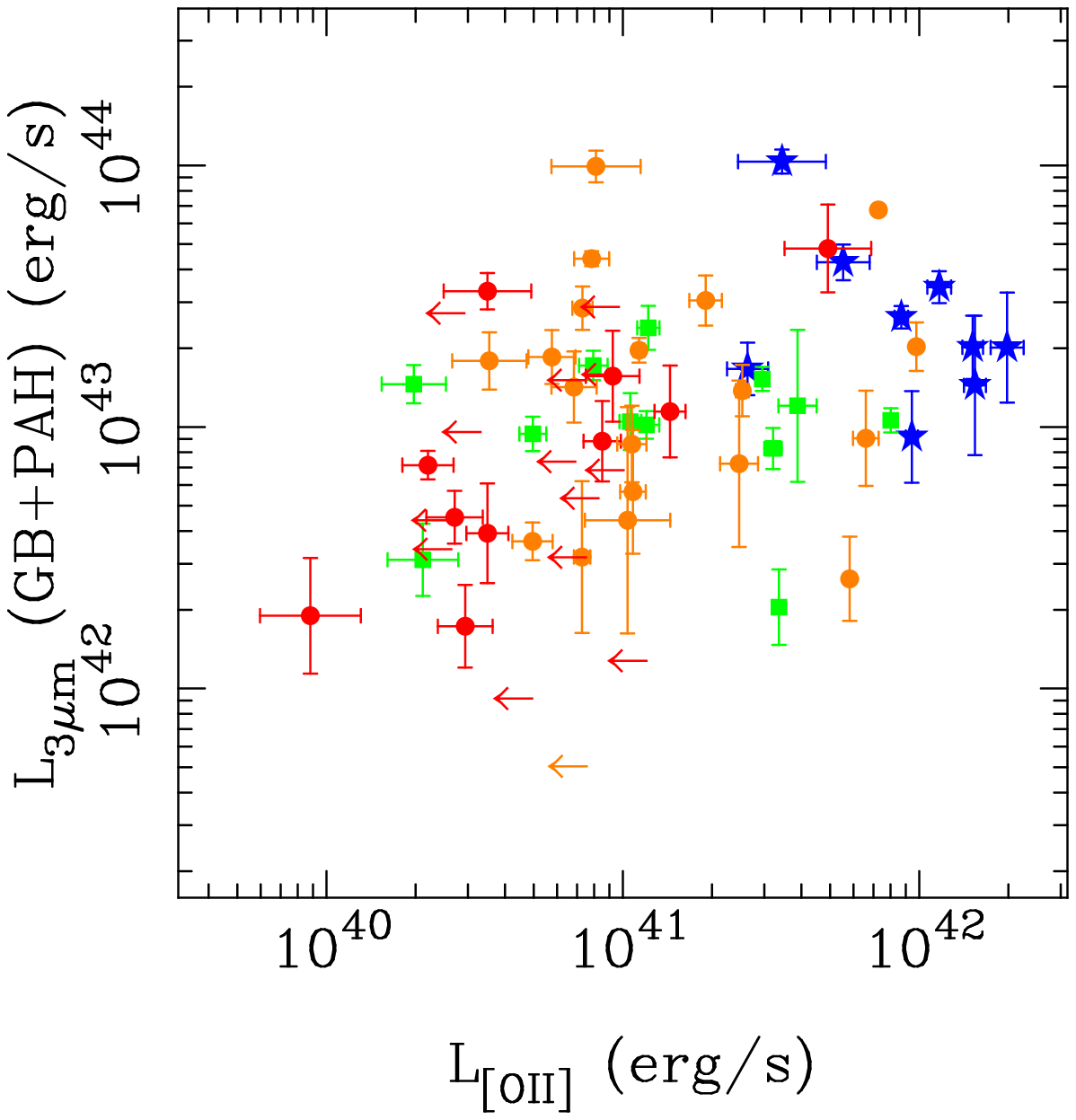}{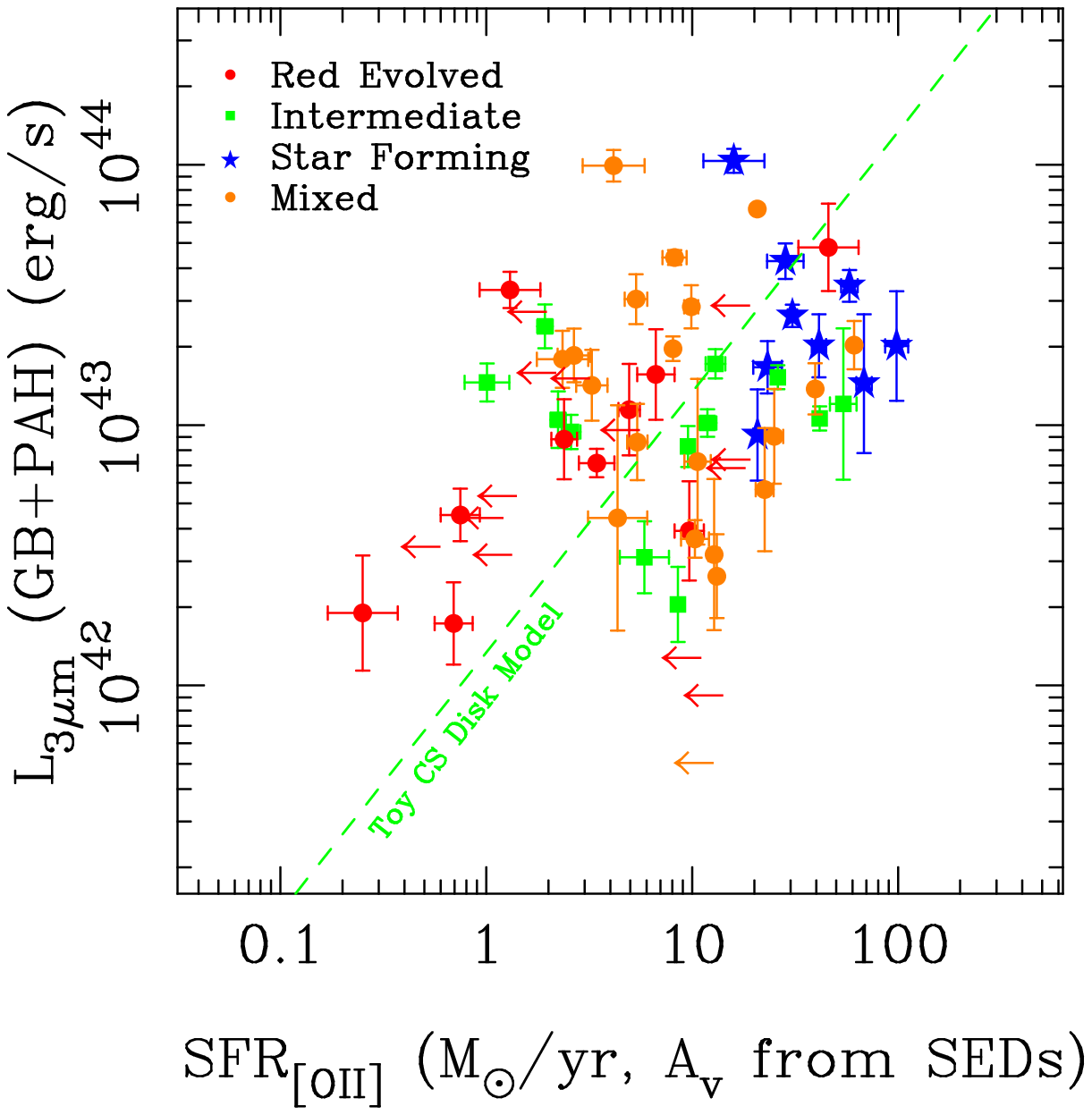}
\caption{Luminosity of the GB+PAH component is correlated ($r=0.49\pm0.06$) to the [\ion{O}{2}] line luminosity (left panel) and the SFR derived from the [\ion{O}{2}] line with dust attenuation correction as described in Section \ref{s:sfr} (right panel). The dashed green line is the luminosity expected from an NIR excess due to circumstellar disks (that last for 1\,Myr) in massive star forming regions, assuming a simple model that scales linearly with the SFR (see Equation (\ref{eq:disk}) in Section \ref{s:disks}). The total luminosity of this model matches the galaxy's total luminosity in the NIR. Galaxies are indicated by spectral types as star-forming (blue stars), evolved (red circles), intermediate (green squares), and mixed (orange circles). (A color version of this figure is available in the online journal.)  }
\label{fig:3micron_vs_sfr}
\end{center}
\end{figure*}

We quantify the amplitude of the NIR excess in our sample as the luminosity in the additional GB+PAH component at 3\,\micron.  The stellar component determined from SED fits is subtracted from this quantity such that both the light from stars formed in the evolved population and the instantaneous burst are excluded. We find that 
this NIR luminosity is correlated ($r=0.49\pm0.06$) with a galaxy's intrinsic [\ion{O}{2}] line luminosity and its corresponding SFR derived from this line as demonstrated in Figure~\ref{fig:3micron_vs_sfr}. SFRs are converted according to \citet{jun05}, except we apply a non-uniform dust attenuation correction using the dust attenuation given as an output parameter in the SED fitting procedure. Because the [\ion{O}{2}] line
probes the formation of massive stars, the correlation indicates that the NIR excess in these galaxies is primarily driven by star formation, 
and is likely powered by the radiation field provided by massive hot stars. 

The correlation between the 3\,\micron~luminosity of the GB+PAH component and the SFR is not strongly dependent on model choices. Both the PEGASE.2 and M05 models yield a similar correlation, although we have chosen to highlight the results from the M05 models which are shown to be a better representation of the NIR stellar light of globular clusters in the Magellanic Clouds due to their treatment of TP-AGB stars. However, some dependence would be expected on the choice of the dust model, in this case the GB+PAH component described in \ref{s:twocomp}, as the broad continuum modeling in the NIR is different in various dust models (see \citealt{des90,dra07} for some examples of theoretical dust models). These dust components may also successfully model the NIR excess, but data at longer infrared wavelengths, combined with NIR and MIR observations, are required to correctly model the behavior of the dust.

\section{Interpretation}
\label{s:dis}

The correlation between the NIR excess emission and SFR
is interesting for a number of reasons. As noted in the previous
section, it provides us with a potentially useful mechanism
for determining the instantaneous SFR from the flux density at rest-frame 2-5\,\micron. Unlike
visible-wavelength indicators of the SFR like the UV continuum or the [\ion{O}{2}] line, the amount of light extinguished 
by dust is likely to be minimal at 3-4\,\micron, 
so that the rest-frame $J-L$ color may be useful as a sensitive and nearly 
extinction-free tracer of star-formation. 

The further importance of the results presented in the
previous section depends on the nature of the sources responsible
for the excess emission. As we have already noted,
the source of the emission may be a component
of a galaxy that at other wavelengths is overwhelmed 
by either starlight or re-radiated emission from large dust grains (note in 
Figure~\ref{fig:sed} that a $\sim1000$\,K blackbody
 peaks at the minimum of these components). If, for example,
it could be shown that the emission originates from 
protostellar or protoplanetary disk emission, then
the excess emission might present us with an opportunity to probe the formation rate 
of planetary systems at high redshifts. On the other hand,
if the emission originates in the interstellar medium (ISM) or post-AGB stars, it presents us with an opportunity to
learn more about the interplay between radiation-driven winds and chemical enrichment
in young galaxies. Furthermore, if low-luminosity AGNs are responsible 
for the excess, the correlation with the SFR indicates a connection
between the growth of black holes and star-formation in our mass-selected sample.

The five best candidates for excess emission are:
\begin{enumerate} 
\item Dust heated from low-luminosity AGNs \citep{dad07,rif09}
\item The high-redshift counterpart to the
interstellar cirrus emission seen in our own Galaxy \citep{ber94,fla06}.
\item Reflection nebulae  \citep{sel83,sel84,sel96}.
\item Post-AGB stars/planetary nebulae \citep{phi05,der06}. 
\item Protostellar/protoplanetary circumstellar 
disks in massive star forming 
regions \citep{hai00,mae05,mae06,lon07}.
\end{enumerate}

In the following sections we investigate whether these
candidates can be responsible for the observed excess. We
first consider large scale, non-stellar, sources of NIR emission
 in Section \ref{s:nonstellar} and then in Section \ref{s:stellar} we extrapolate
  the NIR luminosity from dust emission around stellar candidates, to
  determine if it is enough to account for the total galactic NIR excess emission.

\subsection{Non-stellar candidate sources of excess emission}\label{s:nonstellar}

\subsubsection{Low luminosity AGN candidates}\label{s:agn}

As noted in Section \ref{s:obsIRAC} and \ref{s:mips}, IRAC \& MIPS colors suggest 
that contamination of our sample from strong AGN is likely to be small. However, 
recent MIR spectroscopy of a survey of nearby bright galaxies from \citet{gou09} 
suggests that optical spectroscopic surveys can miss up to half of the AGN population 
due to large dust obscuration. Indeed, NIR and MIR excesses observed 
in a deeper K-selected sample than ours are attributed by \citet{dad07} to 
be due to AGN. They find that 20\%-30\% of their sample show MIR excesses 
higher than those expected from UV derived SFRs alone.
On the other hand, \citet{don08} show that MIR AGN detection criteria, 
although successful at detecting an AGN, can be significantly contaminated 
by star-forming galaxies and the fraction of MIR sources with an AGN is $\sim10\%$ 
at MIR flux densities less than 300\,$\mu Jy$ (only 2/15 of our objects with 
MIPS detections have MIR flux densities greater than 300\,$\mu Jy$). Thus, 
we can rule out strong AGNs as a candidate for the excess emission which
is seen in the majority of our sample.

Since weak AGNs outnumber strong AGNs by a large factor (e.g., \citealt{tru07}), the 
degree of contamination by AGNs in any sample depends on depth and we 
cannot rule out contamination by weak AGNs at the $\sim20\%$ level. 
It is therefore interesting to consider whether hot dust heated by a contaminating 
sample of weak AGN at this level could contribute significantly to the 3\,\micron~  
excess seen in our observations. 

We think this is implausible for a number of reasons. First, heating large 
quantities of hot dust to the required 800\,K -- 1200\,K is likely to be important 
only for strong AGN. For example, \citep{rif09} studied the NIR spectral properties 
of 24 Seyfert galaxies, and they show that the NIR continuum in AGN is 
well-described using a three component model made up of a featureless 
continuum plus two hot blackbody components \citep{rif09}. We can 
certainly exclude a dominant featureless continuum AGN component 
from our sample on the basis of our IRAC colors (see Figure~\ref{fig:IRACcolour}), 
and based on the absence of a significant K-band excess. Figure~\ref{fig:sed} shows an SED for Mrk\,231, a typical AGN, which has excess emission blueward of 2\,\micron, we do not find this type of excess in our photometry (see Figure~\ref{fig:restframeexcess}). These authors 
then show that SyI and SyII samples have markedly different contributions 
from hot dust: 90\% of SyI systems show evidence for hot dust, as opposed to only 
25\% for the SyII population. Therefore, even if our sample is being contaminated 
by a population of weak AGN, only a small subset of these is likely to be contributing 
significantly to the 3\,\micron~ excess. This is clearly inconsisent with 
Figure~\ref{fig:3micron_vs_sfr}, which shows that the excess is a common 
characteristic of star-forming systems, reinforcing the view that star formation, and 
not an AGN, is primarily powering the NIR excess in our K-selected sample.

\subsubsection{Galactic cirrus}

Could the NIR continuum excess seen simply be a scaled up galactic cirrus component?  This idea seems attractive, because after all the canonical SED
being used to model the component is based on da Cunha et al. (2008)'s NIR emission model which is shown to match the observed colors of the galactic cirrus. Attributing the excess to 
cirrus would seem a natural interpretation, provided the energetics can be made to work.

Large-scale observations by \textit{Spitzer} and \textit{COBE} of the galactic cirrus reveal a continuum component in the NIR, whose field to field variations are weak and has a strong contribution to the IRAC [3.6\,\micron] channel, contributing 50\%-80\% of the flux, while the PAH feature at 3.3\,\micron~contributes the remaining flux \citep{fla06} .  Two papers have quoted the irradiance of the galactic cirrus in the $L$-band and they are similar, on average, for two fields at different galactic latitudes. DIRBE measurements from \citet{ber94} find $\nu F_{\nu} = 0.21\times10^{-7}\,\mathrm{W\,m^{-2}\,sr^{-1}}$ for $\lambda = 3.5\,\micron$ for two strips of the galactic plane covering $-6^\circ$ to $-4^\circ$ and $+4^\circ$ to $+6^\circ$ below and above the galactic plane. Using IRAC measurements of the galactic cirrus, at latitudes ranging from $+0^\circ$ to $+32^\circ$, \citet{fla06} find an average intensity of $I_{\nu} = 0.03\,\mathrm{MJy\,sr^{-1}}$ at $\lambda = 3\,\micron$. On the basis of these measurements, it seems that galactic cirrus has a surface brightness on the order of $5\times 10^{-7}$ Jy/arcsec$^2$. In an Euclidean universe surface brightness is conserved with distance,
although in reality $(1+z)^4$ cosmological surface brightness dimming reduces the apparent surface brightness by a factor of around 40 at $z\sim1.5$. The cirrus is expected to have a surface brightness on the order of several tens of nJy/arcsec$^2$ at $z\sim 1.5$. A typical GDDS galaxy subtends an area around 
1 arcsec$^2$ at these redshifts, so the total
flux contributed from cirrus will be around 50~nJy at a rest-wavelength of 3$\mu$m. Inspection of Figures 5--8 makes it clear that this 
corresponds  $<0.1\%$ of the NIR excess. On the other hand, it may be the case that cirrus at high redshifts is far brighter than is the case in the Milky Way.
Cirrus NIR emission is expected to scale linearly with the interstellar radiation field \citep{des90}, so an interstellar radiation field 1000 times that in our own galaxy is needed. Given the modest SFRs in our sample ranging to a maximum of $100\,M_\sun/yr$ (at most 20--50 times that of the Milky Way), we conclude
that it is unlikely that emission from cirrus dominates the NIR excess, though we cannot rule out a contribution from cirrus at the 10\% level.

\subsection{Stellar candidate sources of excess emission}\label{s:stellar}

\begin{figure*}[t]
\begin{center}
\includegraphics[width=6in]{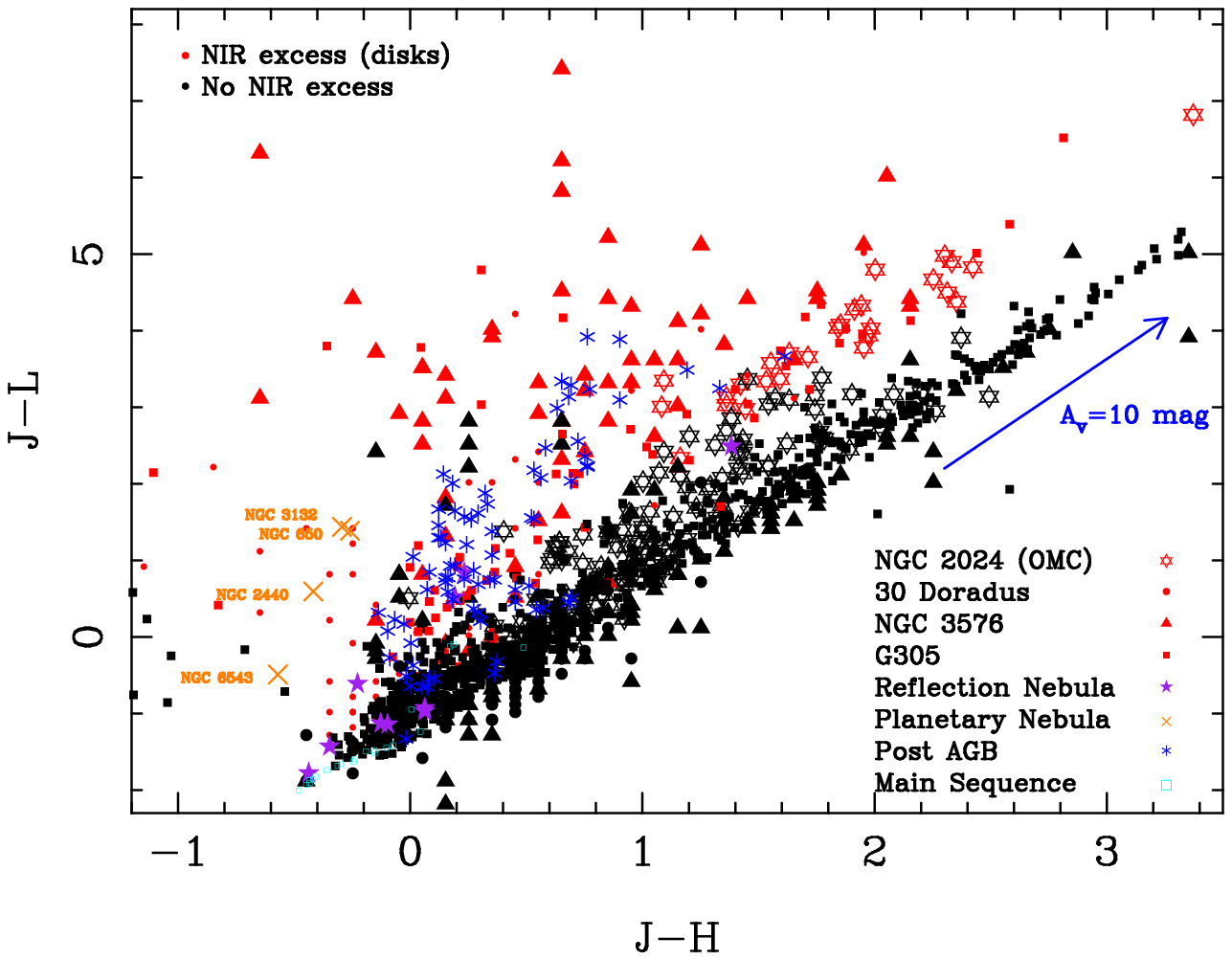}
\caption{$JHL$(AB) color-color plot synthesizing local candidates for the NIR excess. MS standard stars taken from \citet{leg03,leg06} occupy the bottom right of the diagram. Stars embedded in dust will show extinction extending away from the MS along the vector marked in blue.  Stars classified with an infrared excess -- above the excess expected from extinction-- postulated by \citep{mae05,mae06} to be due to circumstellar disks are indicated in red. A range of masses from NGC\,2024 \citep{hai00} in the OMC are plotted as open stars. Massive O-stars in the 30 Dor starburst from \citet{mae05} are plotted as circles as well as O, B stars from starforming regions are plotted from NGC 3576 and G305 \citep{mae06,lon07}. We also plot the NIR photometry of visual reflection nebula from \citet{sel96} (purple stars), planetary nebula (orange crosses) from \citet{hor04} and post-AGB stars (blue asterisks) from  \cite{der06}. (A color version of this figure is available in the online journal.) }
\label{fig:irexcess}
\end{center}
\end{figure*}

In Figure~\ref{fig:irexcess}, we synthesize observations from the Milky Way and the LMC in a $JHL$(AB) color-color diagram for each of the proposed stellar candidates (3-5) for the NIR excess emission. Main-sequence (MS) stars (open cyan squares) not embedded in a dust cloud occupy a region in the bottom left \citep{leg03,leg06}. For dust embedded sources, both the $J-H$ and $J-L$ colors are reddened due to extinction away from the MS along the plotted vector for 10 mag of visual extinction \citep{ind05} .

Using Figure~\ref{fig:irexcess} as a starting point, we now consider in turn
which of these local candidates seems most likely to be responsible for the bulk of the
NIR excess seen in the high-redshift population.
The luminosity in the 
NIR excess component plotted as a function of SFR in Figure~\ref{fig:3micron_vs_sfr}
demonstrates that a useful fiducial luminosity is  $10^{43}$\,erg/s. 
Our simple strategy is to explore how much of each possible candidate 
would be needed to reach this luminosity in a typical galaxy.  

\subsubsection{Reflection nebulae}

NIR excesses are commonly seen in the diffuse emission around visual reflection nebulae (RNae) \citep{sel83,sel96} and  \citet{stu00} show that a scaled spectrum of NGC\,7023, the archetypal RN, matches the spectrum of the starbursting galaxy M82 in the near to mid-IR. We investigate how many RNae are needed to contribute to the NIR excess seen in the GDDS galaxies. Some of the central massive stars powering visual RNae \citep{sel96} also show NIR excess colors as plotted as purple stars in Figure~\ref{fig:irexcess}. For concreteness, we adopt NGC\,7023 (d=430\,pc) as a reference
object, and note that in this object the integrated NIR excess for the star and the nebula are similar in integrated luminosity.  The central star of NGC\,7023, the B2e type star HD\,200755, has an L-band magnitude of $L' = 3.36$ \citep{sel96}. This leads to an intrinsic NIR luminosity of $L_{3.8\,\micron} (\mathrm{star}) = 2.52\times10^{35}\,$erg/s. \citet{sel83} measure the intensity of the diffuse nebular region at a number of positions offset from the central star. The intensity drops off as a function of distance. To estimate the integrated flux, we opt for maximizing the potential contribution of the excess by choosing the strongest intensity measured ($S_{3.8} = 20\times10^{-20}$\,W/m$^2$/Hz$^{-1}$/sr) and integrating the surface brightness over the nebula out to 120\arcsec, where the surface brightness drops by a factor of 10. We find the nebular region to have a maximum integrated luminosity of $L_{3.8\,\micron} (\mathrm{nebula}) = 3.71\times10^{35}\,$erg/s.  

NGC\,7023 is a special RN as the central star also exhibits strong NIR excess in addition to the diffuse NIR excess. However, the amount of integrated NIR luminosity for the diffuse region is not special for this source. NGC\,2023 and NGC\,2068, also relatively bright RNae measured in \cite{sel83}, have similar intrinsic NIR luminosities integrated over their surfaces of $L_{3.8\,\micron} (\mathrm{nebula}) = 8.13\times10^{35}\,$erg/s and $L_{3.8\,\micron} (\mathrm{nebula}) = 2.11\times10^{35}\,$erg/s, respectively. In both cases, we used the distance to the Orion Nebula (d=490\,pc) and assumed the nebula covered a circular aperture in the sky of 180\arcsec~in diameter. To match the fiducial NIR excesses in GDDS galaxies of $\sim10^{43}$\,erg/s, about 10$^{7}$ RNae like NGC\,7023, NGC\,2023 and NGC\,2068 would have to populate an average GDDS galaxy at the time of observation. Locally, RNae typically reside around stars ranging in effective temperatures from 6,800-33,000\,K ($\sim2-20$\,M$_\odot$), but only a fraction of those surveyed (16/23) display NIR excesses \citep{sel83,sel96}. Given that only about $\sim1000$ \citep{mag03} have been observed in our Galaxy, it is unlikely that reflection nebulae are responsible for much of the observed NIR excess, but they cannot be completely ruled out given the selection bias of the RNae surveyed locally.

\subsubsection{Post-AGB stars/planetary nebulae}

NIR excesses with spectral signatures indicative of hot ($\sim1000$\,K) continuum emission have been observed in post-AGB stars \citep{der06} and planetary nebulae \citep{phi05}. Strong mass outflows from intermediate initial mass stars ($4-8\,M_{\odot}$) excites both gas and dust surrounding the luminous (100 - 1000\,L${_\odot}$) post-AGB stars for relatively short timescales of 10$^{4}$\,yrs \citep{der06}. \citet{phi05} attribute $K$-band excesses found in a range of planetary nebula to photon heating of very small grains to temperatures on the order of $800<T_\mathrm{grain} < 1200$\,K. In Figure~\ref{fig:irexcess} we plot the location of several planetary nebulae as orange crosses. In a study of 51 post-AGB stars, \citet{der06} find that all post-AGB stars contain large IR excesses with dust excess starting near the sublimation temperature, irrespective of the effective temperature of the central star. They argue that in all systems, gravitationally bound dusty disks are present. The disks must be puffed-up to cover a large opening angle for the central star and we argue that the disks have some similarity with the passive disks detected around young stellar objects (see Section \ref{s:disks}). The dust excesses of post-AGB stars are noted to be bluer with smaller mid- and far-IR colors compared to the excesses found around Herbig Ae/Be stars and are therefore likely more compact \citep{der06}.

It is straightforward to show that the contribution to the NIR excess from post-AGB stars is not enough to fully account for the NIR excess seen in the integrated luminosity of high-z star-forming galaxies. About $10\%$ of the mass formed in an episode of star formation will be in stars with masses of $4-8\,M_{\odot}$ that will enter the post-AGB phase stage within 200\,Myr after a instantaneous burst of star formation (the average burst age output by our SED modeling is 247\,Myr). To see a noticeable contribution from the short-lived AGB stars to the NIR excess, large-scale and high duty cycle star formation is required. We find in our SED modeling that the average amount of mass formed in an instantaneous burst of star formation is $\log M_{\mathrm{burst}}/M_\odot = 8.5\pm0.9$. In order to reach luminosities of 10$^{43}$erg/s, all of the $4-8\,M_{\odot}$ stars formed in the instantaneous burst would have to be in the post-AGB phase at the time of observation, assuming such a star has a $L_\mathrm{NIR} = 100\,L_{\odot}$, a conservative upper limit. This requires 10$^{8.5}$\,M$_{\odot}$ to be formed in just 10$^{4}$ years \citep{der06}, the typical timescale for the AGB phase and for all the 4-8\,M$_{\odot}$ stars to enter post-MS evolution at the same time, an unrealistic scenario since we know the spread in turnoff ages for the masses are larger than 10$^4$\,years. We also note that due to the delayed response of the post-AGB phase, we would not expect a correlation between NIR excess in AGB stars and \ion{O}{2} luminosity (the SFR indicator traces the most massive population).

\subsubsection{Circumstellar disks around young massive stars }\label{s:disks}

A number of observations (e.g., \citealt{hai00,mae05,mae06,lon07}) suggest that
protostellar/protoplanetary circumstellar disks in star-forming regions are attractive candidates for the
NIR excess. These observations, compiled in Figure~\ref{fig:irexcess}, show that 
the largest L-band excesses locally (up to 5\,mags) are due to excesses
seen in massive stars ($>20\,M_\odot$) in massive star-forming complexes.
Circumstellar disks are proving
ubiquitous in regions of high star-formation, and the inner puffed up rims of circumstellar disks
are predicted to be heated to temperatures high enough to
produce prominent spectral bumps that peak at $2-3\micron$. 

The taxonomy used
to describe such systems has grown rather large, encompassing various classes, stages, 
and groups, and the reader interested in the terminology used to
describe young disks is referred to
the `diskionary' presented by \cite{eva09}. For present purposes it suffices to
consider whether the
IR excess we have detected could be due to any class (or classes) of 
circumstellar disks in a
general sense, 
although for the sake of
concreteness we will base much of our discussion
on the flared disk model presented by \citet{dul01}. 
In this particular
model, the inner part of the disk is removed out to a radius (6\,AU for a B2 star)
of the dust evaporation temperature ($T_{evap} = 1500$\,K). Unlike the top and bottom 
surfaces of the flared disk which receive radiation from the star at a grazing angle,
the inner edge of the disk receives radiation face-on and has a large covering 
fraction of the central star. The result is a disk that is substantially heated to temperatures
up to $T_{evap}$ and the disk becomes puffed up, increasing the surface area of the inner
rim and the NIR emission (see \citealt{dul01} for schematic diagrams).

We once
again refer the reader to Figure~\ref{fig:irexcess}
for a summary of the local observations.
The figure presents a number of data points (open stars) showing the colors of YSOs in NGC\,2024 \citep{hai00}, a young (0.3\,Myr; \citealt{mey96}) massive star-forming region in the Orion molecular cloud (OMC).
\cite{hai00} conclude that $\geq86\%\pm8\%$ of NGC\,2024's members exhibit NIR excesses due to circumstellar disks as indicated by their $JHKL$ colors.
These excesses are seen down to quite low stellar masses, further
suggesting that disks form around the majority of the stars. From their sample of 328 pre-main-sequence 
stars, 45 objects exhibit extremely red excesses of $K - L \geq 1.5$. 
They attribute these to being Class I protostars, but another possible scenario 
we present is that the excess NIR emission could be from disks around the more 
massive objects. 

Moving beyond the OMC to 30 Doradus in the LMC and other large star-forming complexes in our own Galaxy, 
\cite{mae05}, \cite{mae06} and \cite{lon07} have studied IR excess objects using photometry obtained in $JHKL$ and the IRAC bands. These points are also shown in 
Figure~\ref{fig:irexcess}. Objects that have $L$-band IR excesses 
are plotted using red symbols, while those without excesses are plotted as black symbols. The non-IR excess objects roughly follow a linear trend expected due to extinction through the line of sight of the molecular clouds they are embedded in. IR-excess objects in the \cite{mae05} and \cite{mae06} papers show a striking divide from typical reddening to IR-excess with a shift of $\gtrsim1$\,mag redder in L-band. This is likely because they are only reaching the more massive stars in the clusters ($\gtrsim25\,M_\odot$ for 30 Dor and $\gtrsim10\,M_\odot$ for NGC\,3576) and in agreement with the model of \cite{dul01} this would lead to the largest L-band excess. In the OMC sample from \cite{hai00}, which extends down to lower masses, the L-band excess objects are classified based on their NIR colors, however, it is apparent that there is a smooth transition away from the reddening locus, indicative of an NIR excess that increases with stellar mass also in agreement with the disk model of \cite{dul01}. 

Can the NIR excess due to emission from the inner puffed up rims of circumstellar disks provide enough excess flux to account for the excess seen in the integrated light of galaxies? \citet{dul01} calculate the NIR excess fluxes due to disks to be $L_\mathrm{NIR} = (310,8,0.1,0.0035)~L_\odot$, for stars of spectral types B2, A0, G2 and M2, luminosities of $L = (1000,500,1,0.5)\,L_\odot$ and masses of $M = (10,4,1,0.4)\,M_\odot$. Using the results of their Table 1, the NIR excess due to the flared disk is related to the mass of the star by 

\begin{equation}
L_\mathrm{NIR}(M) =  0.0865\,L_\odot \left( \frac{M}{M_\odot} \right) ^{3.48}.
\end{equation}

\noindent We can integrate this over the IMF to figure out the total luminosity excess in the NIR for a given amount of mass formed during a period of star formation. Using the \cite{kro01} IMF, we find the NIR excess due to circumstellar disks is $(L/M)_\mathrm{NIR,disks} = 350\,L_\odot/M_\odot$. While this seems like a high light-to-mass ratio, it is in fact very typical for very young $\lesssim1\,$Myr star-forming regions, which output high IR luminosities.

We can then estimate the total NIR excess luminosity for a galaxy forming stars at a given SFR as:

\begin{equation}\label{eq:disk}
L_{\mathrm{NIR}}(\rm{SFR})_{disk} = 350\,L_\odot / M_\odot \left( \frac{\mathrm{SFR}}{M_\odot/yr} \right) \left( \frac{t_{excess}}{yr} \right).
\end{equation}
\vspace{0.2cm}

\noindent The cluster disk fraction drops drastically from $>80\%$ at $t<1\,$Myr to $<10\%$ by $t\sim5$~Myr \citep{mae05,mae06} so  $t_{excess} = 1\,$Myr is a good order of magnitude estimate for the timescale of circumstellar disks. We plot this simple linear relationship along with the distribution of GDDS data points in right panel of Figure~\ref{fig:3micron_vs_sfr}. The overall normalization of the model 
is remarkably consistent with the data points.  \textit{We find that circumstellar disks with a puffed up inner rim can contribute enough NIR luminosity to be seen over the stellar component of a galaxy and these objects may well be responsible for the NIR excess seen in our high-redshift sample.} 

We emphasize that the model presented is very simple and rather crude. Changing the viewing angle of the disks would lead to changes in the excess, and the relevant timescales may well be shorter than 1\,Myr for the most massive stars which would be provide the bulk of the light. And although we integrate the IMF down to 0.1\,$M_\odot$, the actual contribution from stars with $M<20\,M_\odot$ to the NIR excess luminosity is only $\sim1$\,\% of the total NIR excess luminosity for a \cite{kro01} IMF, so we are really only seeing light from disks around massive young stars. 

Given the simplicity of the assumptions in our toy model,
the order of magnitude agreement with the data seems rather striking. Nevertheless, the scatter in 
the data clearly shows that a simple scaling of the IMF is only an approximation. Any number of physical differences in these galaxies could lead to the scatter from this simple model, such as variations in metalliticity, IMF or reddening. For example, a heavier IMF would result in a higher NIR excess since the fraction of the most massive stars which have the highest excesses would be greater. Alternatively lower metallicity might also lead to a smaller NIR excess for a given galaxy if the density of grains or molecules which cause the excess is smaller. And it is interesting to note that nearby, MIR and PAH emission is shown to be smaller in low-metallicity environments \citep{eng05,mad06}. Another possibility is that the optical measure of the SFR upon which this diagram has been constructed may have been over-estimated, due
to incorrect dust corrections which impact the most actively star-forming galaxies more than they do the quiescent galaxies.
In any case, the important point is that if the IR excess is indeed due to a population of circumstellar/protostellar/protoplanetary disks, modeling
the data shown in Figure~\ref{fig:3micron_vs_sfr} as a function of these input parameters presents us with
an opportunity to determine the rate at which disks form in high-redshift galaxies as a function of 
basic observables, such as a galaxy's global metallicity, mass, and SFR. Furthermore, if the presence of disks around massive young stars can be extrapolated
to lower mass systems which harbor the potential to form planets, then measurements of the NIR excess provide
us with a chance to explore protoplanetary/protostellar disk formation
in a galactic context, opening
up a wide range of interesting studies. The most exciting from a cosmological 
standpoint might be the measurement of the volume-averaged 
cosmic evolution of planet formation.

\section{Conclusions}\label{s:conc}

In a sample of 88 galaxies from the GDDS, a near-infrared excess is detected in the observed IRAC [5.8\,\micron] and [8.0\,\micron] bands. The excess can be modeled as an additional SED component consisting of a modified 850\,K graybody augmented with a mid-IR PAH emission template spectrum \citep{dac08}. The luminosity of the excess SED component is correlated with the SFR of the galaxy, implying the possibility of using NIR luminosity 
as an extinction-free star formation tracer. 
Our exploration of
the origin of the excess flux by examining NIR excesses in our own Galaxy
 has provided a number
of tantalizing possibilities.
From all of the possible NIR excess candidates seen locally, we find 
that circumstellar disks are the most likely candidates for the observed
NIR excess. 
Our simple conversion of a SFR to a total NIR luminosity 
using locally calibrated disk timescales and intrinsic luminosities agree with
 the total integrated NIR luminosity excess measured in the GDDS 
 sample of high-z galaxies. All other candidates require unrealistic 
 abundances or extremely high duty cycle star formation. 
 It seems natural to suppose that the presence of circumstellar 
 disks around massive stars at high redshifts would imply also the 
 presence of disks around less massive stars. These would be systems in which 
 we would expect planets to form. This presents us with an opportunity to probe 
 the formation of planets as seen in their total integrated 
light at high redshifts, at cosmic epochs even before our
own Solar System formed. 

\acknowledgements
We thank the anonymous referee for their thorough review and excellent suggestions which helped improve the quality of this analysis. We also thank Marten van Kerkwijk for useful coffee discussions leading to the interpretations presented in this work. This paper is based on observations obtained at the Gemini Observatory, which is operated by the Association of Universities for Research in Astronomy, Inc., under a cooperative agreement
with the NSF on behalf of the Gemini partnership: the National Science Foundation (United
States), the Science and Technology Facilities Council (United Kingdom), the
National Research Council (Canada), CONICYT (Chile), the Australian Research Council
(Australia), MinistŽrio da Cincia e Tecnologia (Brazil) 
and Ministerio de Ciencia, Tecnolog'a e Innovaci—n Productiva  (Argentina). 
Also, this work is based in part on observations made with the {\it Spitzer Space Telescope}, which is operated by the Jet Propulsion Laboratory, California Institute of Technology under a contract with NASA.
Funding for this research was supported by NSERC, the Government of Ontario and the Canadian Foundation for Innovation. KG acknowledges financial support from Aus-tralian Research Council (ARC) Discovery Project DP0774469. KG also acknowledges support from the David and Lucille Packard Foundation.

\end{document}